\documentclass[aps,prb,twocolumn,groupedaddress,showpacs]{revtex4}
\usepackage{amsfonts}
\usepackage{amssymb}
\usepackage{amsmath}
\usepackage{graphicx}
\usepackage{dcolumn}   
\usepackage{bm}        
\usepackage{flafter}

\begin{document}


\title{Strong electron-phonon coupling in the rare-earth carbide superconductor La$_2$C$_3$}

\date{\today}

\author{J. S. Kim}
\affiliation{Max-Planck-Institut f\"{u}r Festk\"{o}rperforschung,
Heisenbergstra$\beta$e 1, 70569 Stuttgart, Germany}
\author{W.-H. Xie}
\affiliation{Max-Planck-Institut f\"{u}r Festk\"{o}rperforschung,
Heisenbergstra$\beta$e 1, 70569 Stuttgart, Germany}
\author{R. K. Kremer}
\affiliation{Max-Planck-Institut f\"{u}r Festk\"{o}rperforschung,
Heisenbergstra$\beta$e 1, 70569 Stuttgart, Germany}
\author{V. Babizhetskyy}
\affiliation{Max-Planck-Institut f\"{u}r Festk\"{o}rperforschung,
Heisenbergstra$\beta$e 1, 70569 Stuttgart, Germany}
\author{O. Jepsen}
\affiliation{Max-Planck-Institut f\"{u}r Festk\"{o}rperforschung,
Heisenbergstra$\beta$e 1, 70569 Stuttgart, Germany}
\author{A. Simon}
\affiliation{Max-Planck-Institut f\"{u}r Festk\"{o}rperforschung,
Heisenbergstra$\beta$e 1, 70569 Stuttgart, Germany}
\author{K. S. Ahn}
\affiliation{Department of Chemistry, Yonsei University, Wonju
220-710, South Korea}
\author{B. Raquet}
\affiliation{Laboratoire National des Champs Magn\'{e}tiques
Puls\'{e}s, 143 avenue de Rangueil, 31432 Toulouse, France}
\author{H. Rakoto}
\affiliation{Laboratoire National des Champs Magn\'{e}tiques
Puls\'{e}s, 143 avenue de Rangueil, 31432 Toulouse, France}
\author{J.-M. Broto}
\affiliation{Laboratoire National des Champs Magn\'{e}tiques
Puls\'{e}s, 143 avenue de Rangueil, 31432 Toulouse, France}
\author{B. Ouladdiaf}
\affiliation{Institute Laue-Langevin, Grenoble, Cedex 9, France}

\begin{abstract}
We present the results of a crystal structure determination using
neutron powder diffraction as well as the superconducting
properties of the rare-earth sesquicarbide La$_2$C$_3$ ($T_c$
$\approx$ 13.4 K) by means of specific heat and upper critical
field measurements. From the detailed analysis of the specific
heat and a comparison with \textit{ab-initio} electronic structure
calculations, a quantitative estimate of the electron-phonon
coupling strength and the logarithmic average phonon frequency is
made. The electron-phonon coupling constant is determined to
$\lambda_{ph}$ $\sim$ 1.35. The electron-phonon coupling to low
energy phonon modes is found to be the leading mechanism for the
superconductivity. Our results suggest that La$_2$C$_3$ is in the
strong coupling regime, and the relevant phonon modes are
La-related rather than C-C stretching modes. The upper critical
field shows a clear enhancement with respect to the
Werthamer-Helfand-Hohenberg prediction, consistent with strong
electron-phonon coupling. Possible effects on the superconducting
properties due to the non-centrosymmetry of the crystal structure
are discussed.

\end{abstract}

\smallskip


\maketitle

\section{INTRODUCTION}
The recent discovery of superconductivity in
MgB$_2$\cite{MgB2:nagamatsu:syn} and alkaline earth-intercalated
graphites\cite{YbC6:weller:syn} as well as Li at high
pressures\cite{Li:shimizu:pressure,Li:deepa:pressure} has renewed
the interest in electron-phonon ($e$-$ph$) coupled superconductors
without strong electron-electron correlations. The electronic
states and phonon modes relevant for superconductivity vary from
system to system, but in general superconductivity in these
compounds benefits from the light atomic mass of constituents
\textit{e.g.} of boron, carbon and lithium. When a specific part
of the Fermi surface (FS) couples strongly to high frequency
phonon modes, an increase of $T_c$ can be achieved even if the
$e$-$ph$ coupling averaged over the full FS remains moderate. This
mechanism opens a new route to achieve high $T_c$'s by $e$-$ph$
pairing.\cite{eph:picket:review} This is often associated as the
"metallic hydrogen superconductivity"
scenario.\cite{SC:ashcroft:theory}

The rare earth sesquicarbides, $R_2$C$_3$ ($R$ = Rare earths),
which crystallize in the bcc Pu$_2$C$_3$ structure-type, have been
suggested as possible candidates, in which such conditions are
realized. Early on, $T_c$ of La$_2$C$_3$ and Y$_2$C$_3$ was found
to be $\sim$ 11 K, and Th doping in Y$_2$C$_3$ raises $T_c$ to 17
K,\cite{ThY2C3:krupka:syn} comparable with the $T_c$'s of the
$A$15 compounds. In the Pu$_2$C$_3$ structure, C$_2$ dumbbells are
located inside the rare earth metal atom cage. Since the C-C bond
is quite short, the phonon frequency for the C-C stretching phonon
modes is expected to be very high. In fact, recent {\it ab-initio}
calculations showed that the C-C stretching phonon frequency in
Y$_2$C$_3$ is $\sim$1442 cm$^{-1}$.\cite{Y2C3:singh:band}
Therefore $e$-$ph$ coupling between the high frequency phonons and
C-C antibonding states at the Fermi level ($E_{\rm F}$) has been
considered as an origin of the relatively high $T_c$ in Y$_2$C$_3$
and La$_2$C$_3$.

Superconductivity in rare earth sesquicarbides recently regained
attention because of the discovery of 18 K superconductivity in
Y$_2$C$_3$ samples prepared under high pressure ($\sim$ 5
GPa).\cite{Y2C3:amano:syn,Y2C3:nakane:syn} The upper critical
field ($H_{c2}$) is also significantly increased and amounts to
$H_{c2}$(0) $>$30~T.\cite{Y2C3:amano:syn} Electronic structure
calculations\cite{Y2C3:shein:band,Y2C3:singh:band} for Y$_2$C$_3$,
however, demonstrated that the high frequency C-C bond stretching
phonon modes contribute less than 10\% of the total $e$-$ph$
coupling, and rather low frequency Y(-C) phonons must be
considered to be the relevant modes for superconductivity. For
stoichiometric band filling, the $e$-$ph$ coupling constant
$\lambda_{ph}$ is predicted to be $\approx$ 0.6. These results
were consistent with previous $H_{c2}$ studies on La$_2$C$_3$,
which reported $\lambda_{ph}$ $\sim$ 0.8 suggesting moderate
$e$-$ph$ coupling.\cite{La2C3:francavilla:Hc2} However, such a
moderate $e$-$ph$ coupling appears to be too small to generate the
relatively high $T_c$ in the sesquicarbide superconductors.
\cite{Y2C3:singh:band} Therefore, to reconcile with the measured
$T_c$, it is necessary to consider either significant $e$-$ph$
coupling with the high frequency C-C stretching phonon modes or
strong $e$-$ph$ coupling with the low frequency rare earth related
phonon modes. In order to shed light on this controversy, further
experimental studies on the superconducting properties of the
$R_2$C$_3$ are required.

In this paper we focus on one of the rare earth sesquicarbides,
La$_2$C$_3$. We investigate its superconducting properties
together with the crystal and electronic structures. In the rare
earth sesquicarbides, it has been well known that $T_c$ as well as
the structural properties vary significantly depending on the
synthesis and annealing conditions, essentially due to C
deficiency. Therefore in order to carry out reliable studies on
such compounds, one needs to characterize the superconducting
properties simultaneously with the structural properties with
particular attention to C deficiency. Unlike Y$_2$C$_3$, which
requires high pressure and high temperature preparation,
La$_2$C$_3$ can be prepared at ambient pressure conditions using
standard arc melting technique, and larger sample quantities can
easily be synthesized. Recently, we reported that $T_c$ of
La$_2$C$_3$ can be enhanced up to 13.4 K using excess C in the
starting composition of the materials combined with an adequate
post-annealing.\cite{carbide:simon:review,carbide:kremer:review,La2C3:gulden:thesis,La2C3:jskim:syn,La2C3:wang:pressure}

In addition, La$_2$C$_3$ can be a potential system for
investigating the effect of non-centrosymmetry in the structure.
The space group symmetry of $R_2$C$_3$, $I$$\bar{4}$3$d$, belongs
to the tetrahedral crystallographic class $T_d$ lacking a center
of symmetry. When the crystal structure has no center of symmetry
and spin-orbit coupling is significant, the degenerate spin-up and
spin-down bands are mixed and split, which can induce unexpected
superconducting properties. For example, one of the well-known
non-centrosymmetric superconductors
CePt$_3$Si,\cite{CePt3Si:bauer:syn} where Ce $f$-bands possess
significant spin-orbit coupling,\cite{CePt3Si:samokhin:band} shows
unconventional superconducting properties such as high $H_{c2}$(0)
exceeding the Pauli limit\cite{CePt3Si:bauer:syn} and a line node
in the superconducting order
parameter.\cite{CePt3Si:izawa:themalcon,CePt3Si:bonalde:penent}
However, this system also shows a heavy fermion nature, and it is
not clear yet how far the origin of the exotic properties has to
be attributed to the non-centrosymmetry. Thus, it is required to
explore other non-centrosymmetric superconductors $without$ strong
electron correlations.\cite{Li2PtPd3B:yuan:penet} In this respect,
$R_2$C$_3$ compounds with non-magnetic rare earths, can also be
such model compounds to study only the effect of the
non-centrosymmetric structure without the interference from the
magnetism of the constituents.\cite{R2C3:sergienko:nSC} In case of
Y$_2$C$_3$, the spin-orbit coupling for Y $d$-bands may be rather
small, thus we can not expect a significant effect of
non-centrosymmetry. Because of the higher atomic mass of La as
compared to Y, spin-orbit coupling effects in La$_2$C$_3$ is
expected to be more pronounced than in Y$_2$C$_3$. Note that La is
placed next to Ce in the periodic table and accordingly is
expected to generate comparable spin-orbit coupling.

The paper is organized as follows; we first provide experimental
details including a brief description on the synthesis procedures
(Sec. II). Secondly we discuss the crystal structure at low
temperatures gained from neutron powder diffraction (NPD)
investigations (Sec. III) and report the electronic structures of
La$_2$C$_3$ obtained from {\it ab-initio} calculations (Sec. IV).
Specific heat (Sec. V) and the upper critical fields (Sec. VI) on
the samples characterized by NPD are presented. Finally we will
discuss the $e$-$ph$ coupling strength of La$_2$C$_3$ and also the
effects of non-centrosymmetry on the superconducting properties
(Sec. VII).

\section{EXPERIMENTAL}

Polycrystalline samples of La$_2$C$_3$ were synthesized by arc
melting the constituents on a water-cooled Cu crucible under
purified Ar atmosphere. La metal chips (Ames Laboratory, 99.99\%)
were used with spectroscopic grade graphite chips (Deutsche
Carbone, 99.99\%). Before use, the graphite chips were outgased
overnight at 950 $^{\rm o}$C under high vacuum conditions ($P$ $<$
10$^{-5}$ mbar). The starting materials were melted more than 6
times and at each time the button was turned over to ensure
homogeneity. La$_2$C$_3$ is very moisture-sensitive, tending to
decompose readily within a few minutes on exposure to the air.
Consequently all handling of the starting materials and the
samples after synthesis was performed in an Ar filled glove box
(M. Braun $P_{\rm H_2O}$ $<$ 0.1 ppm).

It has been reported that a homogeneity range exists for
La$_2$C$_3$ extending from 45.2\% to 60.2\% atom-\% C
content.\cite{Spedding} After investigation of a series of
La$_{2}$C$_{3-\delta}$ samples,
\cite{La2C3:gulden:thesis,carbide:simon:review,La2C3:jskim:syn} it
was concluded that samples with a carbon deficit upon annealing
phase separate into two superconducting phases with rather sharp
$T_c$'s of $\sim$ 6 K and 13.4 K. The latter transition which can
be attributed to the almost stoichiometric La$_2$C$_3$ phase, is
significantly higher than the $T_c$ of $\sim$ 11 K reported so far
for La$_2$C$_3$. Therefore, to obtain samples with a single sharp
superconducting transition at 13.4 K, it is essential to
compensate possible losses of C in the arc melting and the
subsequent annealing procedure by an excess of carbon up to
10$\%$, although this may result in the impurity phase which is
identified to be LaC$_2$. Heat treatment of the sample buttons was
performed in sealed Ta tubes under purified Ar atmosphere at
1000$^{\rm o}$C. After annealing at high temperatures, the samples
were slowly cooled to room temperature with a rate of 5$^{\rm
o}$C/hour.

Neutron powder diffraction experiments on 2 samples of La$_2$C$_3$
(samples, S1 and S2, $cf$. Table I) were performed at room
temperature using the GEM diffractometer at the ISIS laboratory.
For low temperature neutron powder experiments, another
La$_2$C$_3$ sample (sample S3) of $\approx$ 10 g was sealed into a
vanadium can under 1 bar of He gas and measured down to 5 K on the
D2B diffractometer at the Institut Laue-Langevin (ILL). The
structural parameters as well as the La$_2$C$_3$/LaC$_2$
composition were gained from two-phase Rietveld refinements using
the FULLPROF package.\cite{fullprof} $T_c$ was determined from dc
magnetic susceptibility measurements using a SQUID magnetometer
(Quantum Design, MPMS XL magnetometer). The specific heat ($C_p$)
was measured using a PPMS calorimeter (Quantum Design) employing
the relaxation method. The $C_p$ contribution from the LaC$_2$
impurity phase was estimated from the results of a separate run on
a pure LaC$_2$ sample characterized also with neutron powder
diffraction. To determine the upper critical fields, we measured
the temperature dependence of the resistivity for a bar shape
sample (1\,$\times$\,1 $\times$\,5 mm$^2$) of La$_2$C$_3$ at
different magnetic fields up to 11 T. For magnetic fields higher
than 11 T, we performed magnetization and resistivity measurements
in a pulsed magnetic field up to 30 T at the Laboratoire National
des Champs Magn$\rm \acute{e}$tiques Puls$\rm \acute{e}$s in
Toulouse.

\section{Neutron Powder Diffraction}

\begin{figure}
\includegraphics[width=8.0cm,bb=40 210 560 740]{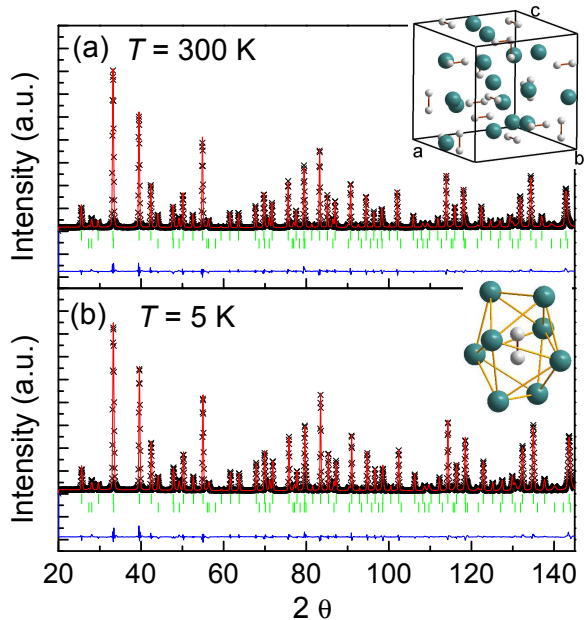}
\caption{\label{npd}(Color online) Neutron powder diffraction
patterns of La$_2$C$_3$ (S3) at (a) 300 K and (b) 5 K (D2B, ILL).
The (red) solid line represents the Rietveld refinement. The
difference between the observed data and the calculated curve is
shown at the bottoms while the Bragg positions are indicated by
upper and lower bars for La$_2$C$_3$ and LaC$_2$, respectively.
The crystal structure of La$_2$C$_3$ and the C-C dumbbell
structure surrounded by the La cage (bisphenoids) are shown in the
insets of (a) and (b), respectively. Large (small) spheres denote
La (C).}
\end{figure}

La$_2$C$_3$ crystallizes with the cubic Pu$_2$C$_3$ structure
(I4$\bar{3}$d) with 8 formula units in the unit cell. As shown in
Fig.~\ref{npd}, C-C dumbbells are located in a distorted
dodecahedral coordination (`bisphenoid') formed by 8 La atoms. So
far, several crystal structure determinations for La$_2$C$_3$ have
been carried out by neutron powder
diffraction\cite{La2C3:atoji1958:neutron,La2C3:atoji1961:neutron}
and also by x-ray single crystal
diffraction.\cite{La2C3:gulden:thesis} There was some variation
found for the lattice parameters ranging from $a$ = 8.804 to 8.818
${\rm \AA}$ at room temperature.
\cite{La2C3:atoji1958:neutron,La2C3:atoji1961:neutron,La2C3:novokshonov:syn}
Such a wide spread of structural parameters has been already known
for Y$_2$C$_3$\cite{Y2C3:amano:syn} as well as other C-containing
superconductors like MgCNi$_3$.\cite{MgCNi3:amos:syn} It is
related to the C content and often affects the superconducting
properties quite substantially. For example, in MgC$_x$Ni$_3$, the
density of state at $E_{\rm F}$, $N(E_{\rm F})$, decreases rather
abruptly due to disorder-induced smearing of the electronic
bands.\cite{MgCNi3:thomas:band} In addition, \textit{e-ph}
coupling mostly due to the low energy Ni-dominated phonon modes is
reduced by C
deficiency.\cite{MgCNi3:johannes:band,MgCxNi3:waelte:Cp} Similar
effects of C deficiency may also cause the large variations of
superconducting properties in $R_2$C$_3$ compounds.
\begin{table*}
\caption{\label{tab:NPD} Structural parameters for La$_2$C$_3$
obtained from Rietveld refinements of neutron powder diffraction
patterns at $T$ = 300 K (S1-S3) and at low temperature down to 5 K
(S3). $a$ is the cubic lattice parameter and $d_{\rm C-C}$ is the
C-C bond length. $u$ and $v$ are the fractional coordinates of the
La 16$c$ site ($u$, $u$, $u$)$_{\rm La}$ and the C 24$d$ sites
($v$, 0, 1/4)$_{\rm C}$, respectively. The reduced $\chi^2$, the
occupancy for C are listed. The employed method, time-of-flight
(TOF) and constant-wavelength (CW) neutron powder diffraction is
also indicated.}
\begin{ruledtabular}
\begin{tabular}{ccccccccc}
Sample & $T$ (K) & $a$ (\AA) &$u$ (La) &$v$ (C) &  $\chi^2$ &$d_{C-C}$ (\AA) & C occupancy  & Method \\
\hline S1 & 300 & 8.80991(8) & 0.05255(3) & 0.30155(9) & 2.074 & 1.2942(16)& 0.983(8) & TOF \\
\hline S2 & 300 & 8.8090(1) & 0.05268(6) & 0.30171(16) & 1.325 & 1.2911(28)& 0.977(5) & TOF \\
\hline S3 & 300 & 8.8096(3) & 0.05255(6) & 0.30144(14) & 8.96 & 1.2977(19)& 0.982(2) & CW  \\
& 200 & 8.8013(3) & 0.05251(5) &  0.30114(14) &  9.42& 1.3001(17)   & & CW \\
& 100 & 8.7936(3) & 0.05251(5) &  0.30085(13) & 10.06& 1.3041(17)  & & CW \\
& 5 & 8.7904(3) & 0.05250(5) &  0.30065(13) & 10.29& 1.3071(17)  & & CW \\
\end{tabular}
\end{ruledtabular}
\end{table*}

Figure \ref{npd} shows the neutron powder diffraction patterns of
La$_2$C$_3$ (S3) collected at $T$ = 300 K and 5 K. In addition to
the reflections which can be ascribed to La$_2$C$_3$, there are
extra weak reflections, which belong to the impurity phase,
identified as LaC$_2$ (I4/mmm). A two-phase Rietveld refinement
was preformed to account for the admixture of the LaC$_2$ phase
whose weight fraction is refined to $\approx$ 18\%. The converged
parameters of the La$_2$C$_3$ phase include the lattice constants,
the fractional coordinates ($u$, $u$, $u$)$_{\rm La}$ of the La
16$c$ site and ($v$, 0, 1/4)$_{\rm C}$ of the C 24$d$ sites, an
isotropic (anisotropic) thermal parameters for the La (C) sites.
After refinements with the aforementioned parameters, in a last
step, the C occupancy was varied. We checked convergence by
varying the C occupancy in the cases where the other refined
parameters are fixed or relaxed. For both cases, a C deficiency of
$\sim$ 2\% is derived consistently. The results of the refinements
for the two patterns collected at $T$ = 300 K - 5 K are listed in
Table \ref{tab:NPD}. Also the results for samples S1 and S2
obtained using time-of-flight (TOF) patterns collected at room
temperature are given.

The C-C distance at room temperature amounts to $\approx$
1.295\,$\rm \AA$, in agreement with the previously reported value,
1.296(9)\,$\rm \AA$ [Ref. \onlinecite{La2C3:atoji1961:neutron}]
but smaller than that reported early on, 1.32\,$\rm
\AA$.\cite{La2C3:atoji1958:neutron} Similar C-C distances have
also been observed in binary rare earth dicarbides and ternary
carbide
halides.\cite{carbide:simon:Handbook,cabide:simon:ZAAC,carbide:henn:Raman,carbide:li:review}
From the studies of electronic behavior on a series of the
dicarbides, $M$C$_2$, the C-C bond length inside the octahedral
metal atom cage is found to be linked to the valence state of the
$M$ atoms.\cite{carbide:li:review} In a simple ionic picture,
CaC$_2$ (Ca$^{2+}$(C$_2$)$^{2-}$) with filled bonding $\pi$ states
but empty antibonding $\pi^*$ and $\sigma^*$ states, has a short
C-C bond distance of $\approx$ 1.2 $\rm \AA$, while with
increasing electron count for the C$_2$ unit $e.g.$ in UC$_2$
(U$^{4+}$(C$_2$)$^{4-}$), the C-C bond distance increases to
$\approx$ 1.35 $\rm \AA$ as the antibonding $\pi^*$ states are
gradually filled. La$_2$C$_3$ has a shorter C-C bond in a
C$_2$$^{4-}$ unit, which is presumably due to the different degree
of charge transfer to antibonding C-C $\pi^*$ states. These
results indicate that the charge transfer between $M$ states and
antibonding $\pi^*$ state of the C-C dimers strongly depends not
only on the valence state of $M$, but also the local environment
around the C-C dimers because of different cage structure.

Interestingly we observed a significant temperature dependence of
the C atom position in contrast to that of the metal atom cage
which contracts in a regular way. With decreasing temperature a
slight increase of the C-C bond distance is found (see Table
\ref{tab:NPD}). A possible explanation of this observation could
be a temperature dependence of the charge transfer between the C-C
$\pi^*$ to the La $d$ states. A carbon-metal orbital overlap will
be affected by volume contraction at low temperatures, which will
modulate the charge transfer between La $d$ and C-C $\pi^*$
states. Another possible origin could be a tilting motion of the
C-C dimers leading to unusual non-elliptical temperature factors
which can not be accounted for by the refinement procedure.
Further studies are ongoing to clarify this issue.

\section{Electronic structure calculations}

The electronic structure of La$_2$C$_3$ has been studied using the
full-potential linear augmented plane wave (LAPW) method with
local orbital extension, within the generalized gradient
approximation (GGA).\cite{wien2k,gga,LAPWdetails} We performed
total-energy calculations for a series of the lattice parameters.
For each lattice parameter the internal parameters (atomic
positions) were relaxed according to the atomic forces. By this
procedure, the fully relaxed structural parameters have been
obtained including the lattice constant as well as the La and C
atomic positions. The calculated equilibrium lattice constant is
$a$ = 8.829 ${\rm \AA}$, with a C-C distance of 1.333\,${\rm \AA}$
and a La-C distance of 2.709\,${\rm \AA}$, corresponding to
internal parameters, $u$$_{\rm La}$ = 0.0524 and $v$$_{\rm C}$ =
0.2995. In comparison with the Rietveld refinement results at
5$\thinspace$K, the calculated lattice parameter is within the
normal error bar of LAPW calculations ($\sim$ 1$\%$), while the
internal atomic parameters for both La and C are in good
agreement.

\begin{figure}
\includegraphics*[width=11cm,angle=-90,bb=0 0 680 465]{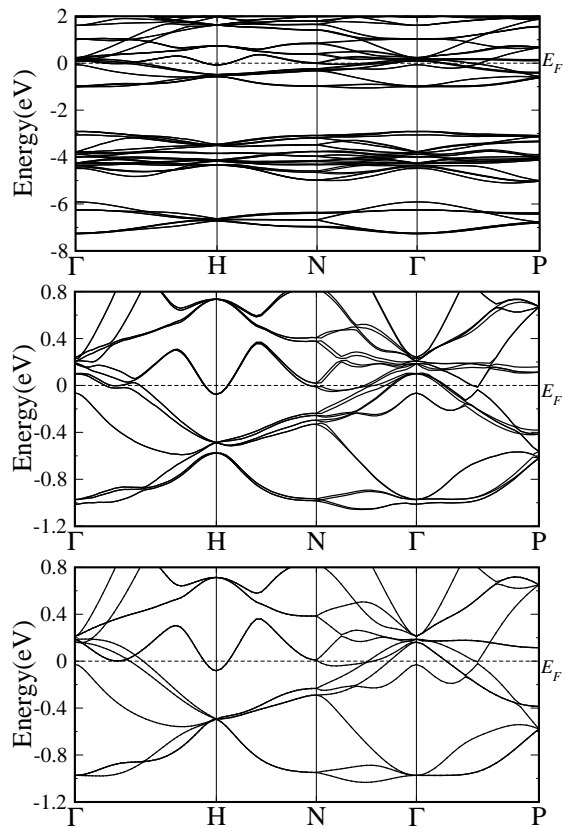}
\caption{\label{bandso} \textit{Upper panel}: Calculated band
structure of La$_2$C$_3$ with spin-orbit coupling, using the
theoretical lattice constant with fully relaxed atomic positions.
For comparison, the band structures near $E_{\rm F}$ with
(\textit{middle panel}) and without (\textit{bottom panel})
spin-orbit coupling are also presented.}
\end{figure}

Figure \ref{bandso} displays the electronic structure with and
without spin-orbit coupling included, using the optimized
structural parameters. There are 6 C-C dimers in the primitive
cell, and the orbitals on these form bonding and antibonding C-C
$\pi$-$\pi^*$ bands. The C 2$s$ derived bands extend from -15.6 eV
to -6.2 eV, relative to the Fermi energy. The lowest bands around
-14 eV and the bands around -7.5 eV are the bonding and
antibonding states, respectively. The bands around -4 eV are the
bonding carbon $p$ bands. The bands crossing the Fermi level,
separated by a 2.5 eV gap from the bonding C-C $\pi$ bands, are
the hybridized La $d$ and C-C antibonding $\pi^*$ states.

\begin{figure}
\includegraphics*[width=5.5cm,angle=-90,bb=0 0 350 490]{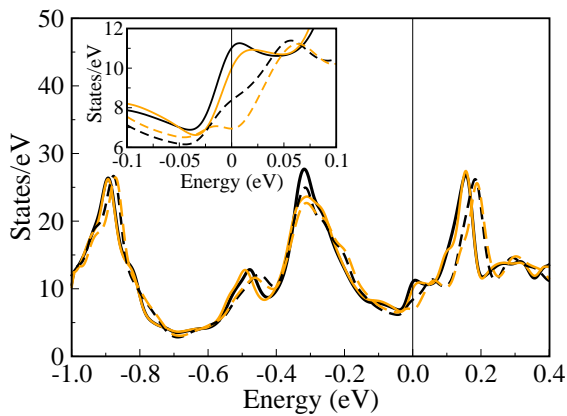}
\caption{(Color online)\label{dos} Electronic density of states of
La$_2$C$_3$ without (black) and with (yellow) spin-orbit coupling
for the stoichiometric (solid) and 2\% C-deficient (dashed)
compounds. The inset shows the DOS near the Fermi level.}
\end{figure}

The electronic structure obtained including spin-orbit coupling is
quite similar to that without spin-orbit coupling, except that the
number of bands is doubled due to the asymmetric spin-orbit
coupling effect (See middle and bottom panels in Fig.
\ref{bandso}). Lacking inversion symmetry in the structure along
with significant spin-orbit coupling, some spin-up and spin-down
bands are mixed and the degeneracy is lifted. Therefore, in
contrast to the previous conjecture,\cite{CePt3Si:sergienko:nSC}
the band splittings due to the asymmetric spin-orbit coupling do
exist, and may also affect the superconducting properties. In
particular, for La$_2$C$_3$, relatively flat bands are crossing
$E_{\rm F}$, thus such a band splitting may alter significantly
the total electronic density of states (DOS) at $E_{\rm F}$. In
Fig. \ref{dos}, we show the DOS with and without spin-orbit
coupling around $E_{\rm F}$. The DOS near the Fermi level is
characterized by a broad minimum centered at about 50 meV below
$E_{\rm F}$. As shown in the inset of Fig. \ref{dos}, for
stoichiometric compounds $N(E_{\rm F})$ is reduced by $\sim$ 10\%
by introducing the spin-orbit coupling.

We also calculated the DOS for the system with 2\% C deficiency by
employing the virtual-crystal approximation. We noticed a shift of
peaks towards higher energies above $E_{\rm F}$. As seen in Fig.
\ref{dos}, while the electronic structure based on the
stoichiometric sample puts $E_{\rm F}$ on the shoulder of the peak
located near the Fermi level, $E_{\rm F}$ for the 2\% C deficient
sample lies close to a minimum. This leads to a reduction of
$N(E_{\rm F})$ by $\sim$ 25\% as compared to the DOS based on the
stoichiometric La$_2$C$_3$ assuming no spin-orbit coupling.
Introducing the spin-orbit coupling, the reduction of $N(E_{\rm
F})$ is even more pronounced and amounts to $\sim$ 30 \%. Thus
slight C deficiency can significantly change the electronic
properties. This provides an explanation of the strong dependence
of $T_c$ and the upper critical field on the composition found in
Y$_2$C$_3$ and
La$_2$C$_3$.\cite{carbide:simon:review,Y2C3:amano:syn,Y2C3:nakane:syn,La2C3:gulden:thesis,carbide:kremer:review,La2C3:jskim:syn}

Singh and Mazin concluded that $\lambda_{ph}$ has a maximum at the
stoichiometric band filling.\cite{Y2C3:singh:band} This is closely
related to the observation that the Fermi energy falls on a peak
in the electronic density of states. Hence, $T_c$ $\approx$ 13.4 K
in the present 2\% C deficient La$_2$C$_3$ samples is already
higher than the previously known $T_c$ $\sim$ 11 K, but it seems
that there is still a possibility to increase $T_c$ further for
stoichiometric La$_2$C$_3$. In this view, the doping dependence of
$T_c$ for La$_2$C$_3$ can also be understood. Th doping in
La$_2$C$_3$ enhances $T_c$ up to 14.3 K,\cite{ThLa2C3:giorgi:syn}
while Lu or Y doping slightly decreases
$T_c$.\cite{La2C3:novokshonov:syn} Considering the electron count
for Th, Th doping at the La sites will donate more electrons than
La, in contrast to Y or Lu substitution. For the previously
investigated La$_2$C$_3$ samples ($T_c$ $\sim$ 11 K), the reduced
electron concentration due to C deficiency, is compensated by Th
doping, but not by Lu or Y doping. Therefore, enhancement of $T_c$
by Th doping can be attributed to the recovery of the band filling
towards the stoichiometric value.

\section{Specific heat}

\begin{figure}
\includegraphics[width=6.5cm,bb=0 220 555 750]{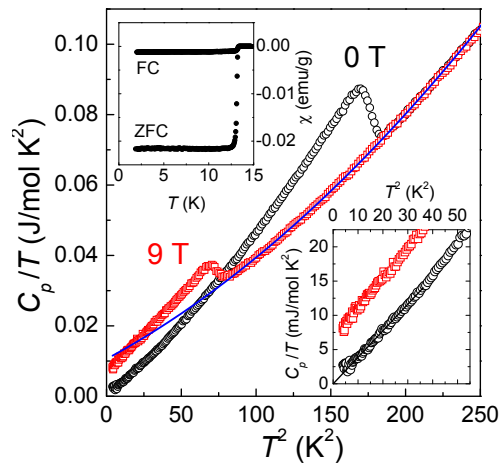}
\caption{(Color online)\label{cp} Temperature dependence of the
specific heat of La$_2$C$_3$ at $H$= 0 and 9 T. The solid (black)
line through the data points at $H$ = 9 T (also in the right
inset) is a fit described in the text. The insets show the low
temperature behavior of $C_p$($T$)/$T$ and the magnetic
susceptibility with zero-field cooled (ZFC) and field-cooled (FC)
modes.}
\end{figure}

Figure \ref{cp} shows the temperature dependence of the specific
heat of La$_2$C$_3$ (S1) at $H$ = 0 and $H$ = 9 T. The
contribution of LaC$_2$ to the total sample capacity, which
amounts to $\lesssim$ 5$\%$ over the whole temperature range, was
subtracted. Note that $T_c$ of LaC$_2$ is $\lesssim$ 1.6
K,\cite{LaC2:ahn:Tc,LaC2:giorgi:Tc} below the temperature range of
the present measurements. There is no offset of $C_p$/$T$ at $H$=
0 as the temperature approaches zero as can be seen in the inset
of Fig. \ref{cp}. This proves that the $C_p$ contribution from the
non-superconducting part, LaC$_2$ has been completely accounted
for the subtraction. A sharp anomaly at $\approx$ 13.4 K is
clearly resolved indicating bulk superconductivity in the
La$_2$C$_3$ sample, which has not been observed
previously.\cite{R2C3:muto:Cp} The onset of 13.4 K determined from
the specific heat jump is consistent with that obtained from the
susceptibility measurements. At $H$ = 9 T, the superconducting
anomaly is shifted to lower temperatures. A clear deviation from a
Debye $T^3$ law is observed in the normal state $C_p$, which can
be attributed to a contribution from low-lying Einstein phonon
modes.\cite{CaC6:jskim:Cp}

\begin{figure}
\includegraphics[width=6.5cm,bb=5 230 530 760]{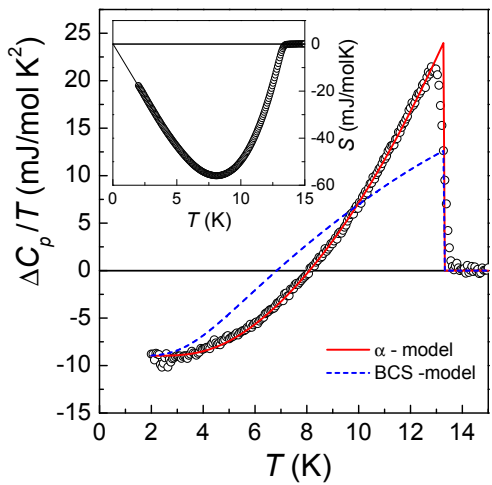}
\caption{(Color online) \label{Csc} Superconducting part of the
electronic specific heat. The (red) solid line is the best fit
according to the $\alpha$-model assuming an isotropic $s$-wave BCS
gap as described in the text. The (blue) dashed line represents
the BCS result. The inset shows the entropy conservation for the
electronic specific heat in the superconducting state.}
\end{figure}
In order to account for the normal state $C_p$, we fitted the data
obtained at $H$ = 9 T by a polynomial,
\begin{equation}
C_p(T) = \gamma_{\rm N} T + C_{lattice} (T) = \gamma_{\rm N} T +
\beta T^3 + \delta T^5.
\end{equation}
Here, $\gamma_{\rm N}$ is the Sommerfeld coefficient and $\beta$
is related to the Debye temperature $\Theta_{\rm D}$(0) via
\begin{equation}
\Theta_{\rm D}^3(0) \thinspace [{\rm K}] = 1944n/\beta \thinspace
[{\rm J/molK^4}],
\end{equation}
where $n$ is the number of atoms per formula unit. Since the
applied magnetic field $H$= 9 T is not sufficient to suppress
superconductivity completely, we used a constraint according
to
\begin{equation} \int_0^{T_c} C_p(T)/T dT = \int_0^{T_c}
(\gamma_{\rm N} + \beta T^2 + \delta T^4) dT,
\end{equation}
to assure the entropy conservation for the superconducting state.

The solid line in Fig. \ref{cp} is the best fit to the $H$\,=\,9 T
data for $T_c$($H$\,=\,9 T) $<$ $T$ $<$ 15 K, yielding the
parameters, $\gamma_{\rm N}$ = 10.60(4) mJ/mol K$^2$, $\beta$ =
228.2(11) $\mu$J/mol K$^4$, and $\delta$ = 0.5988 $\mu$J/mol
K$^6$. The measured $\gamma_{\rm N}$ is much higher and more
realistic than those previously reported from specific
heat\cite{R2C3:muto:Cp} and upper critical fields measurements.
\cite{La2C3:francavilla:Hc2} The corresponding Debye temperature
$\Theta_{\rm D}$(0) = 349(1) K is comparable with that found in
previous work.\cite{R2C3:muto:Cp}

The specific heat difference $\Delta C_p (T)/T$ between the normal
and superconducting state and the entropy for the superconducting
state is shown in Fig.~\ref{Csc}. For comparison, we also plot the
BCS curve for the weak $e$-$ph$ coupling limit. A clear deviation
from the BCS curve is observed. The $C_p$ anomaly at $T_c$ as well
as the intersection temperature where $\Delta C_p$/$T$ = 0 is much
higher than expected from the BCS prediction. The solid (red) line
is the theoretical fit based on the `$\alpha$-model' assuming an
isotropic $s$-wave BCS gap $\Delta$($T$) scaled by the adjustable
parameter, $\alpha$ = $\Delta$(0)/$k_{\rm
B}$$T_c$.\cite{Cp:padamsee:alpha} For the weak coupling limit
$\alpha$ is 1.76. The detailed temperature dependence of $\Delta
C_p$/$T$ was fitted by two adjustable parameters: $\alpha$ and the
Sommerfeld coefficient $\gamma_{\rm N}$. The data are very well
reproduced by $\alpha$ = 2.43 and $\gamma_{\rm N}$ = 8.97 mJ/mol
K$^2$. The $\gamma_{\rm N}$ value from the $\alpha$ model fit is
slightly smaller than obtained from the normal state $C_p$, but in
view of the experimental resolution and the accuracy of the model,
it can be considered as satisfactory agreement. The normalized
specific heat jump, $\Delta$$C_p$/$\gamma_{\rm N}$$T_c$ is also
higher than the weak limit BCS value, 1.426 giving a clear
indication for an enhanced $e$-$ph$ coupling.

\begin{figure}
\includegraphics[width=6.5cm,bb=20 270 520 765]{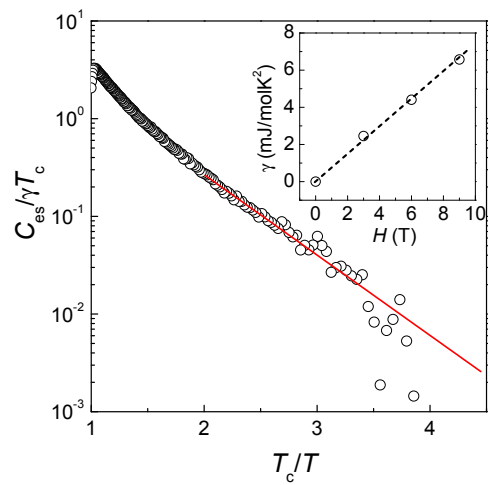}
\caption{(Color online) \label{lowCp}  The electronic contribution
of the specific heat $C_{es}$ is plotted on a logarithmic scale
versus $T_c$/$T$. The (red) straight line is an exponential fit
for 2.0 $<$ $T_c$/$T$ $<$ 4.5. The magnetic field dependence of
Sommerfeld coefficient $\gamma$($H$) is presented in the inset
with the dashed line as a guide to the eye.}
\end{figure}
The normalized electronic specific heat, $C_{es}$/$\gamma_{\rm
N}$$T_c$ in the superconducting state is shown in Fig.
\ref{lowCp}. The solid line is an exponential fit to the data for
2.0 $\leq$ $T_c$/$T$ $\leq$ 4.5 using the form
$C_{es}$/$\gamma_{\rm N}$$T_c$ $\propto$
$\exp$(-0.82$\alpha$$T_c$/$T$) with $\alpha$ =2.31(5). $C_{es}$
exponentially vanishes for $T$ $\rightarrow$ 0 K, clearly
manifesting the absence of gap nodes in the superconducting order
parameter. The $\alpha$ value is somewhat lower than found from
the $\alpha$-model fit in which the ratio $\Delta$(0)/$k_{\rm
B}$$T_c$ is largely determined by the shape of the $C_p$ jump near
$T_c$. However the discrepancy is less than 10$\%$, which is
probably due to a restricted temperature range. The magnetic field
dependence of $\gamma(H)$ at $T$ $\rightarrow$ 0 K is estimated
from the linear fit of $C_p$/$T$ versus $T^2$ for 2 K $\leq$ $T$
$\leq$ 4 K, and it is found to increase linearly with $H$ for low
fields up to $H$ $\approx$ 0.4 $H_{c2}$. This behavior
consistently supports a fully-gapped and almost isotropic
superconducting order parameter.\cite{Cp:nakai:gammaH}

\begin{figure}
\includegraphics[width=6.5cm,bb=10 300 530 760]{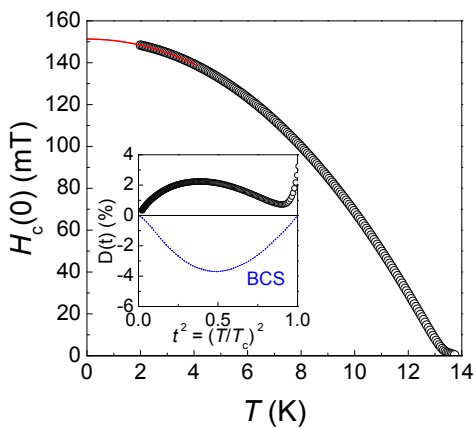}
\caption{(Color online) \label{Hc} The temperature dependence of
the thermodynamic critical field $H_c$($T$) estimated from the
specific heat in the superconducting state. The solid (red) line
is a best fit (see the text). The inset shows the deviation
function, $D$($t$) = $H_c$($T$)/$H_c$(0)-(1-$t^2$) with $t^2$ =
($T$/$T_c$)$^2$. The weak coupling BCS curve is also presented for
comparison.}
\end{figure}

The thermodynamic critical field $H_c$($T$) can be determined
using $H_c$($T$) = $\sqrt{-8\pi\Delta F}$ where $\Delta F$ is the
free energy extracted from the specific heat in the
superconducting state, $\Delta C_p$ = -$T$d$^2$($\Delta
F$)/d$T^2$. The temperature dependence of $H_c$($T$) is shown in
Fig. \ref{Hc}. Based on the BCS theory, $H_c$(0) is derived from a
fit to the temperature dependence of $H_c(T)$ for $T$
$\rightarrow$ 0 which is given by
\begin{equation}
\left(\frac{H_c(T)}{H_c(0)} \right)^2 = 1 - 2.12 \beta \left(
\frac{T}{T_c}\right)^2,
\end{equation}
where the empirical parameter $\beta$ is $\sim$ 1 for the weak
coupling limit but reduced with increasing $e$-$ph$ coupling. The
solid line in Fig.~\ref{Hc} is a best fit yielding $H_c$(0)=
151.3(1) mT and $\beta$= 0.7933(3) indicating as well strong
$e$-$ph$ coupling. The deviation function, $D$($t$) =
$H_c$($T$)/$H_c$(0)-(1-$t^2$) with $t$ = $T$/$T_c$ as well
provides clear signatures for strong $e$-$ph$ coupling. In the
case of strongly $e$-$ph$ coupled superconductors, $D(t)$ is
positive and shows a broad maximum while the weak-coupling BCS
curve exhibits a negative dip at $t^2$ $\approx$ 0.5. As shown in
the inset of Fig. \ref{Hc}, $D$($t$) for La$_2$C$_3$ passes
through a maximum at $t^2$ $\approx$ 0.5.

Based on the specific heat results for the superconducting states,
we conclude that La$_2$C$_3$ is an $s$-wave, single gap
superconductor with strong \textit{e-ph} coupling. Recently,
Harada \emph{et al.} from $^{13}$C nuclear-magnetic-resonance
measurements suggested a multi-gap superconductivity for
Y$_2$C$_3$.\cite{Y2C3:harada:NMR} Since the Fermi surface consists
of several sheets from hybridized La $d$ and C $p$ states, we
cannot rule out such a possibility for La$_2$C$_3$ as well.
However, as shown above, the specific heat for the superconducting
state can be well explained in terms of a single superconducting
gap. Usually, in a system where the $d$ bands are hybridized with
$s$ or $p$ bands, the disparity between the bands is reduced, and
impurity scattering significantly smears out the two-gap
superconductivity.\cite{MgB2:mazin:impurity,MgB2:kortus:impurity}
Considering the polycrystalline nature of the sample, therefore,
it is rather unlikely that, if any, two-gap superconductivity
survives. Further studies are required to clarify the possible
two-gap superconductivity in the sesquicarbide systems.

\section{Upper critical fields}

\begin{figure}
\includegraphics[width=7 cm,bb=10 10 240 300]{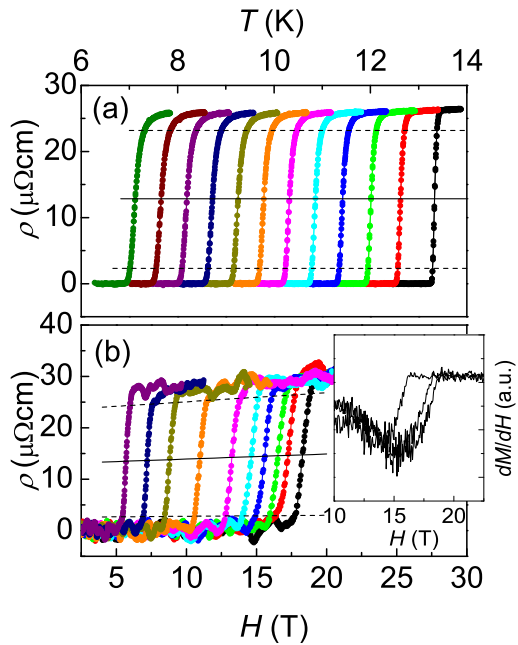}
\caption{\label{MR}(Color online) (a) Temperature dependence of
the resistivity ($\rho(T)$) under magnetic fields from 0 to 11 T
increasing by $\Delta H$ = 1 T, from right to left. (b)
Magnetoresistivity ($\rho(H)$) at different temperatures at $T$ =
10,..., 2.4 K from left to right. The (black) solid line presents
50\% of the resistive transition, while two (black) dashed lines
above and below denote 90\% and 10\% of the resistive transition.
Note that $\rho(H)$ in (b) shows a slight positive
magnetoresistance. The inset of (b) presents the first derivative
of the magnetization versus field curve at $T$ = 4.2, 2.5, and 1.4
K, from left to right.}
\end{figure}
Figure \ref{MR}(a) shows the temperature dependence of the
resistivity under different magnetic fields up to $H$ = 11 T.
$T_c$ is determined at 50\% decrease from the normal state
resistivity value and the transition width is taken as the
temperature interval between 10\% and 90\% of the transition. In
order to determine $H_{c2}$($T$) at low temperatures, the
magnetoresistance was measured up to $\approx$ 30 T at different
temperatures from $T$ = 1.8 to 10 K (Fig.~\ref{MR}(b)). The
transition width remains almost unchanged down to low
temperatures. As a consistency check we also carried out high
field magnetization measurements at low temperatures.
$H_{c2}$($T$) is determined as a kink of the first derivative of
$M(H)$ (See the inset of Fig. \ref{MR}).

\begin{figure}
\includegraphics[width=6.5 cm,bb=5 215 560 740]{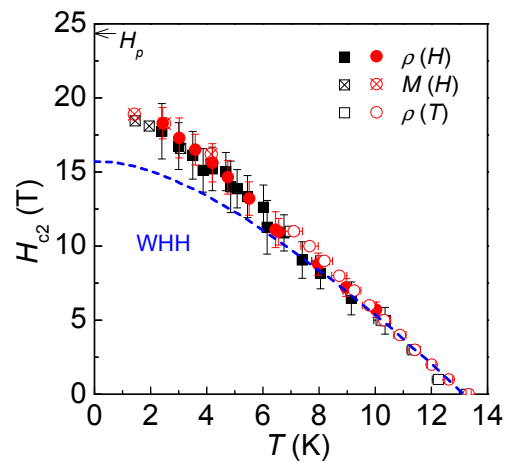}
\caption{\label{Hc2}(Color online) The temperature dependence of
the upper critical field for La$_2$C$_3$ samples, S1 (square) and
S3 (circle), estimated from $\rho$($H$) (filled symbols), $M$($H$)
(crossed symbols), and $\rho$($T$) (open symbols) measurements.
For comparison, the Werthammer-Helfand-Hohenberg curve with the
Maki parameter $\alpha$ = 0 and the spin-orbit scattering
$\lambda_{\rm SO}$ = 0 (see Ref. \onlinecite{WHH}) is presented
with (blue) dashed line. The Pauli limit $H_p$ is indicated by the
arrow.}
\end{figure}

The $H_{c2}(T)$ data obtained from the three different methods are
compiled in Fig. \ref{Hc2}. For comparison, we plotted the
Werthammer-Helfand-Hohenberg (WHH) prediction for conventional
superconductors.\cite{WHH} There is a clear deviation from the WHH
behavior in the $H_{c2}(T)$ curve. This is in contrast to the
previous report on the $H_{c2}$($T$) of La$_2$C$_3$ sample ($T_c$
$\sim$ 11 K), which followed the WHH prediction rather well. Such
an enhancement of $H_{c2}$($T$) has been attributed to several
different origins such as localization effects in highly
disordered superconductors,\cite{hc2:coffey:theory} anisotropy of
the Fermi surface,\cite{hc2:kita:theory} and strong \textit{e-ph}
coupling.\cite{hc2:marsiglio:theory,hc2:bulaevskii:theory} First,
considering disorder effects, the application of the magnetic
field weakens the localization effects and thus reduces the
Coulomb pseudopotential ($\mu^*$). This in turn strengthens the
superconductivity and leads to an enhancement of $H_{c2}$ as the
temperature decreases and larger fields have to be applied to
drive the system into the normal state. Therefore, the enhancement
of $H_{c2}$ is closely linked to the negative magnetoresistance in
the normal state.\cite{hc2:coffey:theory} The positive
magneto-resistance observed in La$_2$C$_3$ (See Fig.~\ref{MR}),
however, indicates that the localization effects tend to be
enhanced under high magnetic fields, which rules out the
possibility of disorder-induced enhancement of $H_{c2}$. Secondly,
when the FS is distorted from the spherical shape, an increase of
$H_{c2}$ becomes more pronounced due to an anisotropy in the
FS.\cite{hc2:kita:theory} Electron structure calculations for
La$_2$C$_3$ unveil rather complex multisheets of the Fermi
surface, similar to that of Y$_2$C$_3$ (\textit{cf.} Fig. 5 in
Ref.~\onlinecite{Y2C3:singh:band}). A quantitative comparison
between experiment and theory is beyond the scope of this work,
but considering the distorted FS shape, the FS anisotropy will
most likely contribute to the enhancement of $H_{c2}$ at low
temperatures. Note, however, that the previous report on
$H_{c2}$($T$) for La$_2$C$_3$ ($T_c$ $\sim$ 11 K) with C
deficiency showed good agreement with the WHH
curve.\cite{La2C3:francavilla:Hc2} With C deficiency, the detailed
FS can be modified, but the overall FS shape would be preserved,
thus similar enhancement of $H_{c2}$ at low temperatures is
expected even for C-deficient samples with lower $T_c$. Therefore
it seems that the FS anisotropy cannot be the only source for
$H_{c2}$ enhancement. Another possibility is strong \textit{e-ph}
coupling. When the \textit{e-ph} coupling is in the strong
coupling regime ($\lambda_{ph}$ $>$ 1), $H_{c2}$($T$) starts to
deviate from the WHH curve and remains linear down to lower
temperatures showing finally an upward curvature for higher
\textit{e-ph} coupling.\cite{hc2:bulaevskii:theory} As shown in
Fig.~\ref{Hc2}, $H_{c2}$($T$) for La$_2$C$_3$ shows an almost
linear temperature dependence down to $\sim$ 0.2$T_c$. From the
specific heat studies, we already found evidence that the
\textit{e-ph} coupling is in the strong coupling regime, thus it
will also affect the temperature dependence of $H_{c2}$ for
La$_2$C$_3$.

From an extrapolation, we obtained $H_{c2}(0)$ $\approx$ 19 T.
Even though $H_{c2}$(0) is much more enhanced than the WHH
prediction, it is still clearly below the Pauli limit of $H_p$(0)
= $\Delta$(0)/$2\sqrt{2}\mu_{\rm B}$ = 1.83 $k_{\rm B}$$T_c$
$\approx$ 24.5 T for the weak coupling limit. Considering the
strong \textit{e-ph} coupling with enhanced $\Delta$(0)/$k_{\rm
B}$$T_c$ = 2.31 - 2.43, $H_p$(0) can be even higher up to $\sim$
32 T. Therefore, $H_{c2}(0)$ $\approx$ 19 T, well below the Pauli
limit, indicates that $H_{c2}$ is mainly determined by orbital
depairing under magnetic fields. This result suggests that the
possible effect due to non-centrosymmetry is not significant for
$H_{c2}$ as we will discuss below (see Sec. VII).

Based on the $H_{c2}(0)$ and $H_c(0)$ values, we estimated the
superconducting parameters for La$_2$C$_3$. According to the
Ginzburg-Landau (GL) theory, the upper critical field $H_{c2}(0)$,
the lower critical field $H_{c1}(0)$ and the thermodynamic
critical field $H_c(0)$ can be described with the GL coherence
length $\xi(0)$ and the GL parameter $\kappa(0)$ =
$\lambda(0)$/$\xi(0)$ according to
\begin{equation}
H_{c2}(0) = \frac{\Phi_0}{2\pi\xi(0)},
\end{equation}
\begin{equation}
H_{c1}(0) = \frac{H_{c}^2(0)}{H_{c2}(0)}[ \ln \kappa(0) +0.08],
\end{equation}
\begin{equation}
H_c(0) = \frac{H_{c2}(0)}{\sqrt{2}\kappa(0)},
\end{equation}
where $\Phi_0$ is the flux quantum. With $H_{c2}(0)$ = 19 T and
$H_c(0)$ = 151 mT, we obtained $\xi(0)$ $\approx$ 42 $\rm \AA$ and
$\lambda (0)$= 373 nm. The GL parameter, $\kappa(0)$ is $\approx$
90, indicating type-II superconductivity. Similar values have also
been reported for the 14 K phase of  Y$_2$C$_3$ ($T_c$ = 13.9 K):
$H_{c1}(0)$ = 3.5 mT, $\lambda(0)$ = 430 nm, $H_{c2}$ = 24.7 T,
and $\xi(0)$ = 36 $\rm \AA$.\cite{Y2C3:akimitsu:Cp}

\section{Discussion}
First we briefly discuss the effect of non-centrosymmetry in
La$_2$C$_3$. There have been several characteristic features
reported concerning the effect of the lack of centrosymmetry on
the superconducting properties : (1) enhancement of $H_{c2}$(0)
above the Pauli limit (2) presence of a line node in the
superconducting gap parameters. Furthermore, it has been suggested
that a description of the superconducting properties should at
least include two bands due to the non-centrosymmetry in rare
earth sesquicarbides.\cite{R2C3:sergienko:nSC} Our electronic
structure calculations confirm the splitting in the electronic
bands near $E_{\rm F}$ indicating that the spin degeneracy is
lifted due to the sizable asymmetric spin-orbit coupling.

\begin{table}
\caption{\label{tab:noncen} Energy splitting of the bands at $E_F$
($\Delta$$E$) and $T_c$ for several non-centrosymmetric
superconductors.}
\begin{ruledtabular}
\begin{tabular}{ccccc}
 & $\Delta$$E$ (meV) & $\Delta$$E$/$k_{\rm B}$$T_c$ &$T_c$ (K)& Ref. \\
\hline
La$_2$C$_3$ &  20 - 30 & $\sim$ 20 & 13.4 & this work \\
CePt$_3$Si & 50 - 200 & $>$ 1000 & 0.75 & \onlinecite{CePt3Si:samokhin:band}  \\
Cd$_2$Re$_2$O$_7$ & $\sim$ 70 & $\sim$ 700 & $\sim$ 1 & \onlinecite{Cd2Re2O7:eguchi:band} \\
Li$_2$Pt$_3$B & $\sim$ 200 & $>$ 1000 & $\sim$ 2 & \onlinecite{Li2R3B:lee:band}  \\
Li$_2$Pd$_3$B & 20 - 50 & $>$ 35 - 85 & $\sim$ 7 &  \onlinecite{Li2R3B:lee:band}\\
\end{tabular}
\end{ruledtabular}
\end{table}

However, even though $H_{c2}$(0) is clearly enhanced, it does not
exceed the paramagnetic limit (Fig. \ref{Hc2}). The specific heat
at low temperatures also reveals that the superconducting gap in
La$_2$C$_3$ has an isotropic $s$-wave symmetry. Therefore the
effect of the non-centrosymmetry appears to be not significant in
La$_2$C$_3$. Note that the band splitting with respect to the
superconducting transition temperature is much smaller for
La$_2$C$_3$ than in other non-centrosymmetric superconductors (See
Table \ref{tab:noncen}). In Li$_2$Pd$_3$B where the band splitting
due to non-centrosymmetry is comparable with that of La$_2$C$_3$,
conventional BCS type behavior with the isotropic superconducting
gap has been found from penetration depth measurements using
muon-spin-rotation experiments.\cite{Li2PdB3:khasanov:muSR}
Therefore for La$_2$C$_3$, the asymmetric spin-orbit coupling
appears to be not strong enough to induce a significant effect on
the superconducting properties. In this respect, it would be very
interesting to study heavier rare earth metal carbides such as
Lu$_2$C$_3$ ($T_c$ $\approx$ 15 K)\cite{Lu2C3:vereshchagin:syn}
and Th$_2$C$_3$ ($T_c$ $\approx$ 4 K).\cite{ThLa2C3:giorgi:syn}

For $e$-$ph$ coupled superconductors,
Carbotte\cite{carbotte:review} has proposed that the
characteristic thermodynamic quantities follow empirical formulas
which can be described by one adjustable parameter, $x$ =
$\omega_{ln}$/$T_c$ where $\omega_{ln}$ is the logarithmic
averaged phonon frequency:
\begin{equation}
\frac{2\Delta(0)}{k_{\rm B}T_c} = 3.53 \left[ 1 + 12.5 x^{-2} {\rm
ln} \frac{x}{2}\right],
\end{equation}
\begin{equation}
\frac{\Delta C_p(T_c)}{\gamma_{\rm N}T_c} = 1.43 \left[ 1 + 53
x^{-2} {\rm ln} \frac{x}{3}\right],
\end{equation}
\begin{equation}
\frac{\Delta C_p(T)-\Delta C_p(T_c)}{\gamma_{\rm N}T_c-\gamma_{\rm
N}T} = -3.77 \left[ 1 + 117 x^{-2} {\rm ln} \frac{x}{2.9}\right],
\end{equation}
\begin{equation}
\frac{\gamma_{\rm N} T_c^2}{H_c^2(0)} = 0.168 \left[ 1 - 12.2
x^{-2} {\rm ln} \frac{x}{3}\right],
\end{equation}
\begin{equation}
\frac{H_c(0)}{dH_c(T)/dT|_{T_c}T_c} = 0.576 \left[ 1 - 13.4 x^{-2}
{\rm ln} \frac{x}{3.5}\right].
\end{equation}
This analysis has been successfully applied to various metal-alloy
superconductors,\cite{carbotte:review} and also to other recently
discovered carbon-contained superconductors such as borocarbides
\cite{manalo:Cp,michor:Cp} and MgCNi$_3$.\cite{MgCNi3:waelt:Cp}

\begin{figure}
\includegraphics[width=6.5 cm,bb=35 25 545 770]{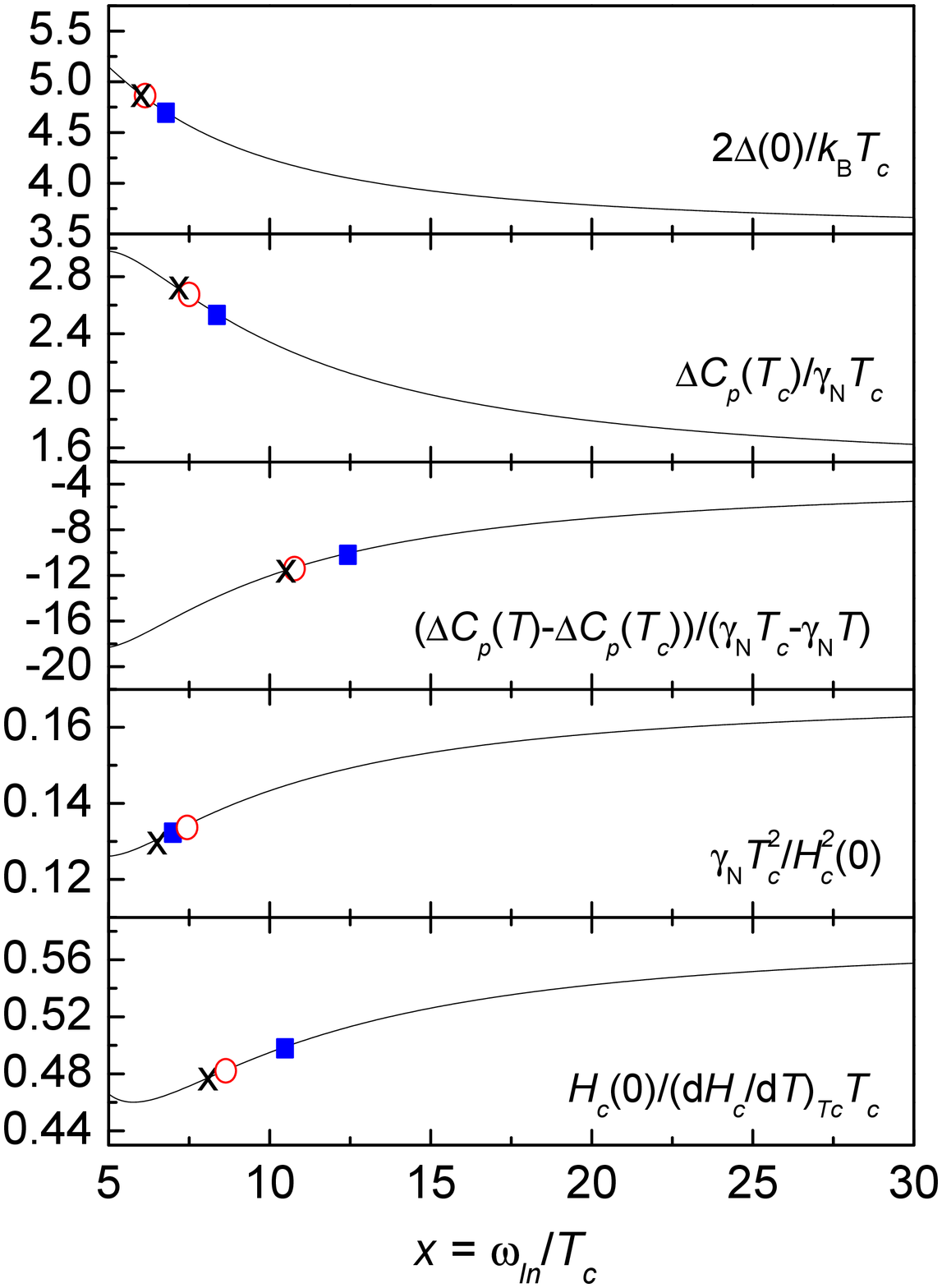}
\caption{(Color online) \label{wln} Several thermodynamic
quantities as a function of the logarithmic averaged phonon
frequency (solid line) according to Eqs. (8)-(12). The
thermodynamic quantities estimated from the specific heat data for
the samples are indicated with (red) open circles for S1, (black)
crosses for S2 and (blue) filled squares for S3.}
\end{figure}

Figure \ref{wln} shows the thermodynamic quantities with variation
of $x$ = $\omega_{ln}$/$T_c$ according to Eqs. (8)-(12). The
measured values extracted from the $C_p$ data are also plotted
onto the empirical curves. Each thermodynamic quantity provide a
value for $x$ = $\omega_{ln}$/$T_c$ as indicated for three
La$_2$C$_3$ samples (S1-S3) in Fig. \ref{wln}. From five $x$'s and
$T_c$ = 13.4 K, we obtain a mean value $\omega_{ln}$ $=$ 109 K
$\pm$ 24 K, 101 K $\pm$ 24 K, and 121 K $\pm$ 32 K for the sample
S1, S2, and S3, respectively. Averaging all $\omega_{ln}$'s for
the samples we conclude on $\omega_{ln}$ = 110 K $\pm$ 27 K. This
$\omega_{ln}$ value is only $\sim$ 30\% of the Debye temperature,
$\omega_{\rm D}$ $\approx$ 350 K. The Debye frequency is
determined only from the phonon DOS profile, $F(\omega)$, while
$\omega_{ln}$ results from a weighting by the $e$-$ph$ coupling
function $\alpha^2$($\omega$). Such a reduced $\omega_{ln}$ value
compared to $\omega_{\rm D}$ indicates the importance of the low
energy phonon modes for the superconductivity. The deviation of
the normal state $C_p$ from the Debye $T^3$ law (See.
Fig~\ref{cp}) supports the presence of the low-lying Einstein
phonon modes.

\begin{figure}
\includegraphics[width=6.5 cm,bb=35 250 550 750]{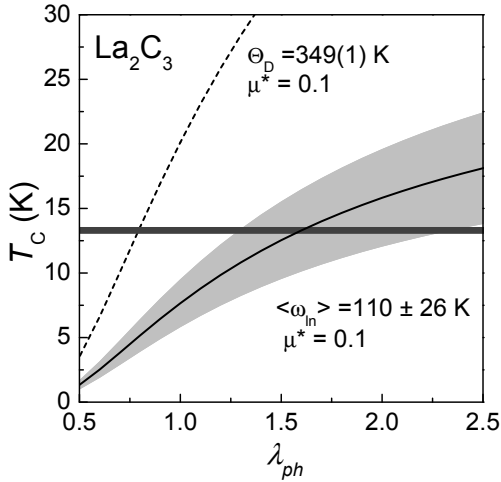}
\caption{\label{lambda} The calculated $T_c$ from the modified
McMillan formula (solid line) as a function of the $e$-$ph$
coupling constant, $\lambda_{ph}$. The borders of the shaded
(grey) areas represent the results obtained for the upper and
lower bounds of $\omega_{ln}$. For comparison, we also plot the
calculated $T_c$ from the original McMillan formula with dashed
line. The horizontal (dark grey) solid line denotes the measured
$T_c$.}
\end{figure}

From {\it ab-initio} calculations for Y$_2$C$_3$, the frequencies
of the full symmetry Raman modes are estimated to be 175 cm$^{-1}$
for a Y dominated mode and 1442 cm$^{-1}$ for an almost pure C-C
bond stretching mode.\cite{Y2C3:singh:band} The relevant phonon
modes for the superconductivity, therefore, are the La dominated
phonon modes rather than the high frequency C-C bond stretching
modes. From simple conversion based on the atomic mass difference
between La and Y, we can estimate the frequency of the La
dominated Raman mode to be $\sim$ 140 cm$^{-1}$. In comparison
with $\omega_{ln}$ $\approx$ 76 cm$^{-1}$ $\pm$ 15 cm$^{-1}$, this
phonon frequency is somewhat higher. However, since the lattice
constant of La$_2$C$_3$ is larger than that of Y$_2$C$_3$, the
interatomic forces are expected to be somewhat weaker for
La$_2$C$_3$. The estimated frequency of the La dominated phonon
mode, therefore, is rather overestimated, and it can be reduced
and come closer to the $\omega_{ln}$.  Recent {\it ab-initio}
calculations for Y$_2$C$_3$ also reported that most of the
$e$-$ph$ coupling emerges from the low frequency Y related phonon
modes, not from the high frequency C-C phonon
modes.\cite{Y2C3:singh:band}

Other relevant phonon modes for superconductivity in La$_2$C$_3$
could be the tilting vibrations of the C-C dimer. For $A$C$_2$
($A$ = Ca, Sr, and Ba) containing the C-C dimers inside the
octahedral metal atom cage, it has been shown that the binding of
the C-C dimer to the surrounding octahedral metal atom cage is
weak, thus allowing a structural transition from tetragonal to a
cubic phase at high temperatures as well as several modifications
of the C-C dimer structure at low
temperatures.\cite{CaC2:knapp:syn} Raman scattering studies on
rare-earth carbide halides also showed that the frequency of the
C-C tilting modes is $\sim$ 400 cm$^{-1}$, much lower than the
stretching modes with a frequency of $\sim$ 1590
cm$^{-1}$.\cite{carbide:henn:Raman} Compared to the octahedral
metal atom cages, the size of a bisphenoid La cage is larger, thus
the binding of the C-C dimer to the La cage could be even weaker.
Consequently the tilting modes of the C-C dimer in La$_2$C$_3$ are
expected to have lower frequencies, which can allow sizable
contribution to $e$-$ph$ coupling. Further studies on the C-C
dimer tilting modes \textit{e.g.} Raman scattering experiments or
\textit{ab-initio} calculations are highly desirable.

Finally we estimate the \textit{e-ph} coupling constant for
La$_2$C$_3$. Based on the Sommerfeld coefficient and the
electronic structure calculation, we obtain $\lambda_{ph}$ from
the determination of $\gamma_{\rm N}$ using the equation
$\gamma_{\rm N}$ = (2$\pi^2$$k_{\rm B}^2$/3)$N$($E_{\rm
F}$)(1+$\lambda_{ph}$). From the DOS calculations for the 2\% C
deficient compounds, we take $N$($E_{\rm F}$)= 1.75 states/eV f.u.
Using $\gamma_{\rm N}$ =10.6 mJ/mol\,K$^2$, the estimated
$\lambda_{ph}$ is 1.4 indicating that the La$_2$C$_3$ clearly
belongs to the strong coupling regime.

In Fig.~\ref{lambda} we plot the calculated $T_c$ from the
modified McMillan formula,
\begin{equation}
T_c =\frac{\omega_{ln}}{1.2} \exp \left[
\frac{-1.04(1+\lambda_{ph})}{\lambda_{ph}-(1+0.62\lambda_{ph})\mu^*}
\right],
\end{equation}
with the measured $T_c$ = 13.4 K (grey horizontal line). For
comparison we also plot the calculated $T_c$ from the original
McMillan formula.\cite{note} In both cases, the Coulomb
pseudopotential $\mu^*$ is fixed to 0.1. With the original
McMillan formula assuming the effective phonon frequency to be the
same as the Debye frequency, we found that the measured $T_c$ is
reproduced with weak \textit{e-ph} coupling, $\lambda_{ph}$ $\sim$
0.75. However such a weak coupling cannot reconcile with the
results of specific heat and upper critical fields as discussed
above. If we use the logarithmic averaged phonon frequency
$\omega_{ln}$ in the modified McMillan formula, Eq. (13), a large
value of $\lambda_{ph}$ is necessary to reproduce the measured
$T_c$. Because of the relatively large error for the calculated
$T_c$ from the modified McMillan formula due to the reduced
accuracy for determining $\omega_{ln}$, it is difficult to
determine the $\lambda_{ph}$ value precisely. However it is
obvious that $\lambda_{ph}$ is larger than 1.3, which is in good
agreement with the estimate from the comparison of the Sommerfeld
coefficient.

In conclusion, specific heat and upper critical field studies were
performed on the sesquicarbide superconductor La$_2$C$_3$ together
with neutron powder diffraction as well as electronic structure
calculations. The main conclusions are the followings. (i) The
density of states near the $E_{\rm F}$ is very sensitive to a
small C deficiency in the sample, which provides an explanation
for the wide scatter of $T_c$'s and superconducting properties
observed in the previous reports. (ii) The temperature and
magnetic field dependence of the specific heat is consistent with
a single gap $s$-wave BCS superconductor. (iii) The logarithmic
averaged phonon frequency for the superconductivity is quite low
suggesting the importance of low energy phonon modes for the
superconductivity. (iv) Even though the band splitting due to the
non-centrosymmetry in the structure and the spin-orbit coupling is
clearly confirmed by the electronic structure calculations, its
effect seems not sizable enough to cause exotic superconducting
properties as observed in other non-centrosymmetric
superconductors. To conclude, all of the summarized features
suggest that La$_2$C$_3$ is a strongly coupled BCS-type
superconductor with an isotropic $s$-wave superconducting gap.

\acknowledgments

The authors acknowledge stimulating discussions with A. Das and O.
Dolgov. We thank E. Br\"ucher and G. Siegle for expert
experimental assistance. Part of this work has been supported by
EuroMagNET under the EU contract RII3-CT-2004-506239.


\begin{thebibliography}{}
\bibitem{LAPWdetails}
In the calculations, we set RK$_{\rm max}$ = 5.5, $l$ = 12 and
G$_{\rm max}$ = 18. The muffin-tin radii of 2.5\thinspace$a_0$ and
1.14\thinspace$a_0$ were chosen for La and C, respectively. For
the spin-orbit calculation, the states up to 10 Ry are included in
the second-variation diagonalization of thespin-orbit Hamiltonian.
\bibitem{note}
The only difference of the original McMillan formular from Eq.
(13) is that the coefficient $\omega_{ln}$/1.2 is replaced by
$\Theta_{\rm D}(0)$/1.45. See Ref. \onlinecite{Tc:allen:theory}.
\end{thebibliography}
\end{document}